\documentclass[preprint,showkeys,aps,prb]{revtex4}
\usepackage[dvips]{graphicx}

\begin{document}

\title{ A Tutorial: Adaptive Runge-Kutta Integration for Stiff Systems: \\
Comparing the Nos\'e and Nos\'e-Hoover Oscillator Dynamics}

\author{
William Graham Hoover and Julien Clinton Sprott and Carol Griswold Hoover                   \\
Ruby Valley Research Institute and Department of Physics University of Wisconsin-Madison    \\
Corresponding Author email : hooverwilliam@yahoo.com                                        \\
Madison, Wisconsin 53706                 \\
Highway Contract 60, Box 601, Ruby Valley, Nevada 89833, USA ;                              \\
}

\date{\today}

\vspace{0.1cm}

\begin{abstract}
``Stiff'' differential equations are commonplace in engineering and dynamical systems.
To solve them we need flexible integrators that can deal with rapidly-changing righthand
sides. This tutorial describes the application of ``adaptive'' [ variable timestep ]
integrators to ``stiff'' mechanical problems encountered in modern applications of
Gibbs' 1902 statistical mechanics. Linear harmonic oscillators subject to
{\it nonlinear} thermal constraints can exhibit either stiff or smooth dynamics.  Two
closely-related examples, Nos\'e's 1984 dynamics and Nos\'e-Hoover 1985 dynamics, are
both based on Hamiltonian mechanics, as was ultimately clarified by Dettmann and Morriss
in 1996. Both these dynamics are consistent with Gibbs' canonical ensemble.  Nos\'e's
dynamics is ``stiff'' and can present severe numerical difficulties. Nos\'e-Hoover
dynamics, though it follows {\it exactly the same} trajectory, is ``smooth'' and
relatively trouble-free.  Our tutorial emphasises the power of {\it adaptive}
integrators to resolve stiff problems like the Nos\'e oscillator. The solutions obtained
illustrate the power of computer graphics to enrich numerical solutions.  Adaptive
integration with computer graphics is basic to an understanding of dynamical systems
and statistical mechanics.  These tools lead naturally into the visualization of
intricate fractal structures formed by chaos as well as elaborate knots tied by regular
nonchaotic dynamics.  This work was invited by the American Journal of Physics.
\end{abstract}


\maketitle

\section{Introduction and Overview}

Nos\'e's very original 1984 work extended classical {\it isoenergetic} Newtonian
molecular dynamics to include a new temperature-based dynamics.  His goal was to 
replicate Gibbs' {\it isothermal} canonical ensemble directly from dynamics. 
Nos\'e based his work on isoenergetic Hamiltonian mechanics\cite{b1,b2,b3,b4,b5,b6}.
The resulting motion equations are typically ``stiff'' and hard to solve. Nos\'e
introduced the idea of ``time-scaling'' to cope with these difficulties.  Hoover
pointed out that a smoothed and improved set of motion equations could be based on 
Liouville's Theorem without the need for time-scaling or a Hamiltonian basis\cite{b4}.
Dettmann furnished a Hamiltonian basis linking both sets of motion equations
a decade later in 1996\cite{b5,b6}.

Both the original Nos\'e dynamics and the improved Nos\'e-Hoover dynamics share
a common flaw.  They are not necessarily``ergodic''.  Their phase-space flows cover
only a fraction of the available states, even for the simplest (harmonic-oscillator)
applications\cite{b5,b6,b7,b8,b9,b10}.  It was not until 2015 that an ergodic analog of
Nos\'e's dynamics was discovered\cite{b9}, reaching a goal of his original
1984 project.

Nos\'e's work provides many opportunities for further study and improvements.  We
include here side-by-side calculations for the (stiff) Nos\'e and (smooth)
Nos\'e-Hoover oscillators.  These two oscillator types follow {\it exactly the same}
phase-space trajectory but at different rates. The dynamics for these oscillators
are related to each other by ``time scaling''. In fact, the
{\it four}-dimensional descriptions of the two models' detailed trajectories are
precisely identical ! But of the two only Nos\'e's oscillator problem exhibits the stiffness
which is the focus of the present work.

The Nos\'e oscillator was designed to model a harmonic oscillator at thermal
equilibrium, characterized by a temperature $T$.  The dynamics Nos\'e developed
has both regular [ stable to small perturbations ] and chaotic [ unstable to
perturbations ] solutions, coexisting in a constant-energy three-dimensional
volume within a four-dimensional phase space.  Within that space there are
infinitely-many regular solutions (concentric tori and stable periodic orbits)
in addition to a single chaotic sea which stretches to infinity and occupies
about six percent of the stationary measure defined by Liouville's Theorem.
Quite remarkably that stationary measure has a simple analytic form.  It is a
smooth three-dimensional Gaussian distribution. Though the simpler regular
stable tori and the periodic orbits they enclose are relatively easy to
solve, motion in Nos\'e's version of the chaotic sea is sufficiently
stiff to require the special solution methods we describe in what follows.

Typical fixed-timestep integrators like leapfrog and fourth-order Runge-Kutta are
ineffective in Nos\'e's chaotic sea.  We emphasize here a family of useful alternatives,
simple variable-timestep ``adaptive'' algorithms\cite{b11}.  We explore Nos\'e's work
with their aid.  {\it Adaptive} techniques vary the numerical algorithms' timestep to
compensate for time-dependent changes in the stiffness of the underlying ordinary
differential equations.  The present Tutorial is intended to introduce students and
researchers to both the Nos\'e and Nos\'e-Hoover oscillator example problems and to
their solution using adaptive integration.  We connect ``stiffness'' with time-stepping,
time-scaling, and Lyapunov instability.  In exploring these numerical features of
our models we come upon intricate topologies with their roots in simple quadratic
differential equations.

We recommend the Nos\'e oscillator to students as a testbed for integrators, computer
graphics, and numerical methods.  This picturesque model provides challenges in
visualizing the fascinating topology of knots and island chains in readily accessible
three- and four-dimensional phase spaces.  The model's simple structure makes it an ideal
introduction to dynamical-systems research.

In Section II we detail the statistical-mechanical background which links the Nos\'e
and Nos\'e-Hoover models.  In Section III we introduce a family of flexible integrators
capable of accurate solutions of the stiff oscillator problems.  In Section IV we choose
and apply a common initial condition for our dynamic investigations of time-scaling and
stiffness.  This choice makes our work reproducible.  We illustrate the evolution of
Nos\'e's time-scaling factor ``$s$'', the adaptive timestep $dt$, and the local Lyapunov
exponent, $\lambda_1(t)$ .  We compare simulations with both double- and quadruple-precision
arithmetic, 64 and 128 binary bits respectively.  In Section V we take stock of what we have
learned and suggest areas for further investigations using our new tools.

\begin{figure}
\includegraphics[width=4.5in,angle=+90.]{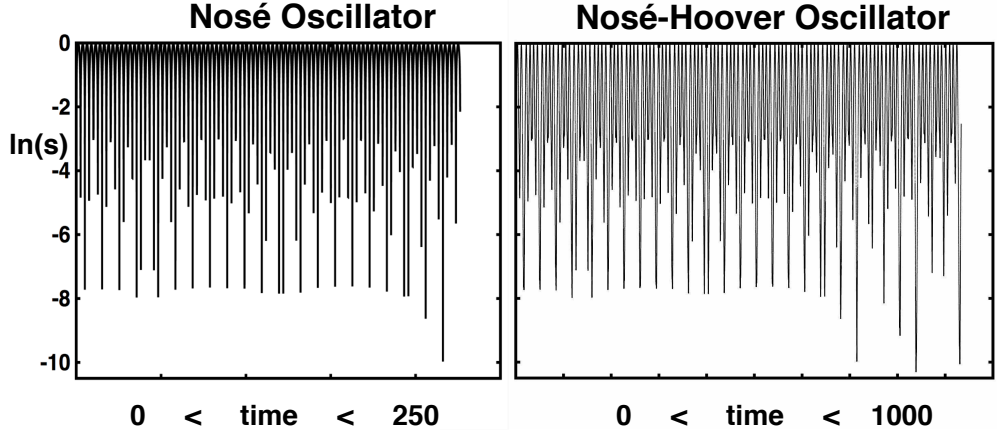}
\caption{
The variation of the [ Nos\'e ] (at the left) and [ Nos\'e-Hoover \#1 ] (at the right)
time-scaling factors $s$ with time.  See subsection IIA of the text for the differential
equations solved here. In principle exactly the same values of $(q,p,s,\zeta)$ and, in
the same order, apply to both models.  Both models were solved for 100 000
adaptive-integrator timesteps.  The adaptive RK4 timestep is doubled whenever the rms
discrepancy between a single $dt$ timestep and two $(dt/2)$ half timesteps is less than
$10^{-12}$ .  The timestep is halved if the discrepancy is greater than $10^{-10}$ .
The minimum value of $s$ within this fragment is about $e^{-10} \simeq 0.00005$ .  The
initial conditions for both these simulations are
$(q,p,s,\zeta,{\cal H}) = (2.4,0,e^{-2.88},0,0)$ , chosen so that the Hamiltonian
${\cal H}$ vanishes. All four Nos\'e-Hoover rates are smaller than the Nos\'e rates
by a factor of $s(t)$ .
}
\end{figure}

\section{Nos\'e and Nos\'e-Hoover Dynamics for Harmonic Oscillators}

In 1984 Shuichi Nos\'e imagined a temperature-dependent Hamiltonian ${\cal H}_N$
with a ``time-scaling'' control variable $0 <s<1$ . This ``invention'', or better
yet, ``discovery'', provides a unique $s$-dependent dynamics.  The speed at which a
thermal trajectory evolves is governed by $s$ , which in turn controls the kinetic
temperature $T$. The dynamics is consistent ( in the ``necessary but insufficient''
sense ) with Gibbs' canonical ensembles of constant-temperature states rather than
the more usual ``microcanonical'' ensembles of constant-energy states. In
conventional oscillator mechanics the total energy ${\cal H} = (1/2)(q^2 + p^2)$ is
constant where  $q$ and $p$ are the oscillator displacement and momentum.
That usual oscillator dynamics generates sine and cosine solutions with vibrational
amplitudes proportional to the square root of the kinetic temperature $p^2$ .
$$
\langle \ q^2 \ \rangle = \langle\ p^2 \ \rangle = T
= \langle\ {\cal H} \ \rangle \ .
$$

In Nos\'e's $s$-dependent mechanics temperature is likewise a measure of the
kinetic energy, $T \propto (p/s)^2$ , where $(p/s)$ is a {\it scaled} Cartesian
momentum component. In the simplest harmonic-oscillator case Nos\'e's approach
uses $s$ to {\it scale} the momentum over a broad range.  The scale factor $s$
varies from its maximum, unity, to less than $ 10^{-9}$ over a billion-timestep
adaptive simulation where the mean timestep is about 0.002.  See the shorter
100 000-timestep oscillator histories shown in {\bf Figure 1}. We detail a useful
approach to such problems in what follows.

Throughout this Tutorial we adopt the simplest possible notation so as to focus on
the basic ideas. We choose to explore the behavior of a harmonic oscillator with
unit mass and force constant.  In keeping with simplicity ( and without loss of
generality ) we also set Boltzmann's constant and the temperature equal to unity.

Nos\'e's {\it time-scaled} Hamiltonian ( where $s$ is the time-scaling factor and
$\zeta$ is its conjugate momentum ) for the isothermal one-dimensional harmonic
oscillator is : 
$$
2{\cal H}_N = q^2 + (p/s)^2 + \ln (s^2) + \zeta^2 \equiv 0 \
[ \ {\rm Nos\acute{e}} \ ] \ .
$$
Here and in what follows  we consistently choose the soon-to-be-explained
significant value  of the Hamiltonian ${\cal H}_N \equiv 0$ . This Hamiltonian
governs the evolution of the four time-dependent variables $(q,p,s,\zeta)$ .
$q$ and $p$ are still the usual oscillator coordinate and momentum, though with
the usual link between velocity and momentum changed.  The Cartesian relation
$\dot q = p$ is replaced by $\dot q = (\partial{\cal H}_N/\partial p) = (p/s^2)$ .
Because the canonical distribution includes states with {\it all} energies,
$0 <{\cal H} < \infty$ , Nos\'e had the idea to include the ``time-scale factor''
$s$, along with its conjugate momentum  $\zeta$, {\it in} the Hamiltonian,
making it possible for the ``scaled'' momentum $(p/s)$ to cover the
{\it infinite} range required by the canonical ensemble's distribution function,
$$
f(q,p)_{\rm canonical} \equiv e^{-q^2/2} e^{-p^2/2}/(2\pi) \ .
$$

Generally Hamilton's motion equations for {\it any} $(q,p)$ coordinate-momentum pair are :
$$
\dot q = +(\partial {\cal H}/\partial p) \ ; \
\dot p = -(\partial {\cal H}/\partial q) \ .
$$
In addition to the $(q,p)$ pair the Nos\'e oscillator has also the
time-scaling variable $s$ and its conjugate momentum $p_s = \zeta$ making up a
second coordinate-momentum pair $(s,\zeta)$.  The motion of the Nos\'e
oscillator in its four-dimensional phase space $(q,p,s,\zeta)$ follows
from his Hamiltonian :
$$
{\cal H}_N \longrightarrow \dot q = (p/s^2) \ ; \ \dot p = -q \ ;
\ \dot s = \zeta \ ; \ \dot \zeta = (p^2/s^3) - (1/s) \
[ \ {\rm \ Nos\acute{e} \ } \ ] \ .
$$
Because the value of the Hamiltonian is constant the motion takes place in the
three-dimensional volume where ${\cal H}_N = 0$ . That volume is unbounded. 
So long as the {\it scaled} kinetic temperature is less than unity, $(p/s)^2 < 1$ ,
{\it any} $(q,\zeta)$ combination is accessible by choosing a sufficiently small
value of $s$ (with an even smaller value of $p < s$) .  We saw in {\bf Figure 1}
that the scale factor $s$ ranges over more than four orders of magnitude.  As a
result Nos\'e's equations of motion present numerical challenges. Fortunately there
is a way to temper the singular behavior of $\dot \zeta$ , as was clarified by
Dettmann and Morriss a dozen years after Nos\'e's work\cite{b6}.

\subsection{Dettmann and Morriss' 1996 Contribution : a New Oscillator Hamiltonian}

Their contribution is an alternative fully-Hamiltonian description of the
{\it same} trajectories but with all of Nos\'e's rates given above multiplied
by $s$ .  Here ${\cal H}_D$ is that Hamiltonian, followed by the new, generally
slower smoother rates it generates :
$$
{\cal H}_D = s{\cal H}_N = 
(1/2)[ \ sq^2 + (p^2/s) + s\ln(s^2) + s\zeta^2 \ ] \equiv 0 \longrightarrow  
$$
$$
\dot q = (p/s) \ ; \ \dot p = -qs \ ;                                
\ \dot s = \zeta s \ ; \ \dot \zeta = (p^2/s^2) - 1 \      
[ \ {\rm \ Nos\acute{e}-Hoover \ \# 1 \ } \ ] \ .
$$
Multiplying Nos\'e's Hamiltonian ( and thus the four rates ) by $s$ when $s$ is
small tames the singular behavior of Nos\'e's mechanics and is equivalent
to a close relative of Nos\'e-Hoover mechanics, ``Nos\'e-Hoover \#1''.  This
scaled-time improvement can be simplified further to get the usual
``Nos\'e-Hoover \#2'' motion equations. Just replace the scaled momentum $(p/s)$
with $p$ :
$$
\dot q = p \ ; \ \dot p = -q - \zeta p \ ; \ \dot s = s\zeta \ ; \ \dot \zeta =
 p^2 - 1 \ [ \ {\rm \ Nos\acute{e}-Hoover \ \# 2 \ } \ ] \ .
$$
From the numerical standpoint an advantage of this second \#2 set is the
{\it irrelevance} of the scaling variable $s$ .  The evolution of $\{ \ q,p,\zeta \ \}$
, still in a three-dimensional space extending to infinity, can be determined
without any consideration of $s$ provided that {\cal H} vanishes.

We urge the reader to focus on the unusual condition ${\cal H}_D \equiv 0$ . This
choice is necessary to the derivation of the [ Nos\'e-Hoover \#1 ] equations in the
absence of time scaling. The vanishing Hamiltonian makes it possible to simplify
the expression for $\dot \zeta$ :
$$
\dot \zeta = -(\partial {\cal H}_D/\partial s) = -(1/2)[ \ q^2 - (p/s)^2
+\ln(s^2) + \zeta^2 + 2 \ ] \stackrel{{\cal H}_D \ = \ 0}{=} (p/s)^2 - 1 \ .
$$

\subsection{Hoover's 1985 Contribution : Nos\'e-Hoover Motion equations {\it via} Liouville}

Let us take a moment to detail Hoover's much simpler 1985 derivation of the three-equation
subset of the $[ \ {\rm \ Nos\acute{e}-Hoover \ \# 2 \ } \ ]$ equations\cite{b4}.  Begin
with the augmented set of oscillator motion equations with a friction coefficient
$\zeta$ which acts in a characteristic relaxation time $\tau$ :
$$
\dot q = p \ ; \ \dot p = -q -\zeta p \ ; \ \dot \zeta = [ \ (p^2/T) - 1 \ ]/\tau^2  \ .
$$
Here $\zeta$ is the ``friction coefficient'' or ``control variable''.  If the kinetic
temperature $p^2$ exceeds the target temperature $T$ the friction increases,
slowing $p$.  If instead $p^2$ is too cool, less than $T$, the friction is reduced
and can become negative, accelerating the oscillator.  Provided only that a stationary
state results, the long-time-averaged value $\langle \ \dot \zeta \ \rangle$ is
necessarily zero so that the kinetic temperature $p^2$ eventually reaches its target :
$$
\langle \ \dot \zeta \ \rangle =
\langle \  [ \ (p^2/T) - 1 \ ]/\tau^2  \ \rangle = 0 \longrightarrow
\langle \ p^2 \ \rangle = T \ .
$$
A remarkable feature of the motion equations is that they leave Gibbs' canonical
distribution function ( or probability density ) unchanged.  Suppose that
$$
f(q,p,\zeta,T) = (\tau/T)(2\pi)^{-3/2}e^{-q^2/2T}e^{-p^2/2T}e^{-(\zeta\tau)^2/2}
$$
and consider the rate of change of probability density in $r=(q,p,\zeta)$ space as
a result of the continuity  equation for the flow $v=(\dot q,\dot p,\dot \zeta)$ :
$$
(\partial f/\partial t) = -\nabla_r \cdot (fv) \equiv
$$
$$
-f[ \ (\partial \dot q/\partial q) + (\partial \dot p/\partial p) +
(\partial \dot \zeta/\partial \zeta) \ ] - \dot q(\partial f/\partial q) 
- \dot p(\partial f/\partial p) - \dot \zeta(\partial f/\partial \zeta) = 
$$
$$
-f[ \ 0 - \zeta + 0 - (q/T)p - (p/T)(-q - \zeta p) -
\zeta\tau^2 \{ \ [ \ (p^2/T) - 1 \ ]/\tau^2 \ \} \ ] \equiv 0 \ .
$$
The vanishing rate of change throughout $(q,p,\zeta)$ space implies that
Gibbs' distribution is left unchanged by the flow.  By design the control
variable responds to $(p^2/T) - 1$ in a characteristic time $\tau$ .

Here and throughout we have consistently modified Nos\'e's original
work\cite{b1,b2}, replacing his $(2/s)$ by $(1/s)$ and his $p_s$ by $\zeta$
in order to match the alternative Nos\'e-Hoover description  of the
oscillator trajectory.  The wide-ranging values of the time-scaling
variable $s$ ( see again {\bf Figure 1} ) make the Nos\'e  equations
so ``stiff'' that an accurate fourth-order fixed-timestep Runge-Kutta
solution of the equations for a typical time of $10^6$ would require
$10^{13}$ timesteps with $dt = 10^{-7}$ .  A few years of laptop
time !

Let us summarize what we know from this Gibbsian and Hamiltonian background
information in statistical mechanics and Nos\'e's temperature-dependent
dynamics.  [1] Nos\'e's mechanics generates stiff motion equations which are
difficult to solve with conventional methods. [2] By scaling the time and
redefining the momentum, $(p/s) \rightarrow p$, a more manageable set of
equations consistent with Gibbs' canonical distribution results.  [3]
Exactly these same motion equations can be derived directly from the
phase-space continuity equation if one insists that the friction
coefficient $\zeta$ imposes the kinetic temperature $T$ on
the dynamics. The situation is ideal because we have several ways to check
our work.  Apart from the time the stiff Nos\'e equations and the smooth
Nos\'e-Hoover equations have identical solutions !  Because our goal
is learning to solve the stiff set :
$$
\dot q = (p/s^2) \ ; \ \dot p = - q \ ; \ \dot s = \zeta \ ; \
\dot \zeta = (p^2/s^3) - (1/s) \ , 
$$
we turn next to developing suitable adaptive integrators.

\section{Stiff Oscillator Solutions {\it via} Adaptive Runge-Kutta}

The original work on the stiff oscillator problem\cite{b3,b4} was frustrated by
huge and rapid variations in $p$ and $s$.  Stiff equations were unusual in the
molecular dynamics simulations of the 1980s.  Researchers with backgrounds in
simulation were not familiar with ``adaptive'' integrators.  On the other hand researchers
in control theory and heat transfer {\it often} used the 1969 Runge-Kutta-Fehlberg
integrator to solve their own stiff equations.  That Fehlberg modification of the
classic Runge-Kutta methods of the early 20th century, compares fourth-order and
fifth-order versions of a trajectory integration for a single step $dt$ .
The two integrals over $dt$ provide a criterion for increasing or decreasing $dt$
on the next timestep.  If the discrepancy is ``too large'' the timestep is reduced.
If it is ``too small'' the timestep is increased.  In this way the discrepancy
between the two estimates can be restricted to an error band.  If the discrepancy is
too large an alternative to proceeding with the better RK5 estimate is to repeat the
current step with smaller and smaller $dt$ until the discrepancy falls within the
acceptable band.  Suitable bandwidths for double precision and quadruple precision
are $10^{-12}$ to $10^{-10}$ and  $10^{-24}$ to $10^{-20}$.

Rather than comparing RK4 and RK5 one can just as well compare an iteration with
$dt$ to two successive iterations with $(dt/2)$ .  We have adopted that choice
here. We compare two RK4 integrations [ $dt {\it \ versus \ } (dt/2)+(dt/2)$ ] , in
these demonstration problems, comparing a full timestep integration of the vector
$x = (q,p,\zeta) {\rm \ or \ } (q,p,s,\zeta)$ from $t$ to $t+dt$ with the result
of  two half-timestep integrations of the initially identical vector $y$ :\\
{\tt
      call rk4(x,xp,dt/2.0d00)\\
      call rk4(x,xp,dt/2.0d00)\\
      call rk4(y,yp,dt/1.0d00)\\
      errerr = (x(1)-y(1))**2 + (x(2)-y(2))**2 + (x(3)-y(3))**2 + (x(4)-y(4))**2\\
      error = dsqrt(errerr)\\
      if(error.gt.10.0d00**(-10)) dt = 0.5d00*dt\\
      if(error.lt.10.0d00**(-12)) dt = 2.0d00*dt\\
}
The variables {\tt (x,xp,y,yp)} are vectors containing the integration variables
{\tt (x,y)} and the righthand sides {\tt (xp,yp)} of the differential equations of
motion.  Comparing the single-timestep {\tt y} solution to the double-timestep
{\tt x} solution provides criteria for increasing or decreasing the timestep $dt$ .
In either case the less-accurate  vector, $y$, is set equal to the more nearly
accurate vector $x$ before the next timestep is undertaken.

\begin{figure}
\includegraphics[width=3.5in,angle=90.]{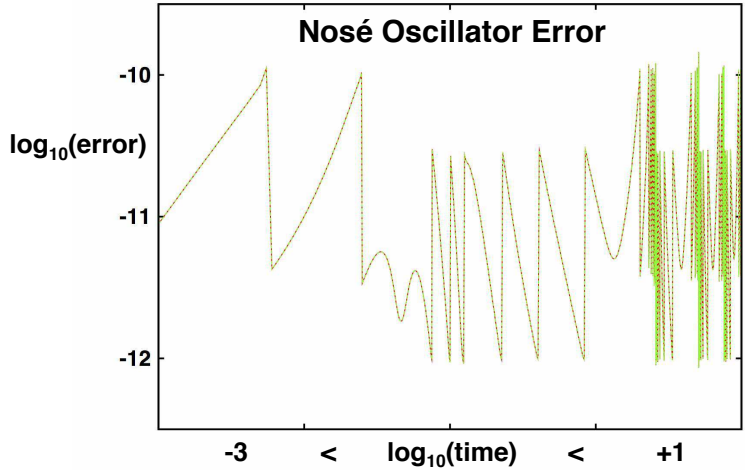}
\caption{
Sample variation of the rms integration errors within the band
from $10^{-12}$ to $10^{-10}$ with $0.001<{\rm time}<10$ .  Our standard initial
condition $(q,p,s,\zeta,{\cal H}) = (2.4,0,e^{-2.88},0,0)$, for which ${\cal H}$
vanishes, was used.  Although both double-precision and quadruple-precision
results are shown here the difference is barely visible on the scale of this plot.
}
\end{figure}

As a quick demonstration problem we choose for our standard initial condition
$(q,p,s,\zeta,{\cal H}) = 2.4,0,e^{-2.88},0,0)$ because this choice corresponds
to the Dettmann Hamiltonian value for which the Nos\'e and Nos\'e-Hoover motion
equations provide ( apart from numerical errors ) {\it identical} trajectories:
$$
2{\cal H}_D = q^2 + (p/s)^2 + \ln(s^2) + \zeta^2 = 2.4^2 + 0 -5.76 + 0 \equiv 0 \ .
$$
{\bf Figure 2} shows a short history of the rms discrepancy between the single
and double-step versions of
$$
\dot q = (p/s^2) \ ; \ \dot p = - q \ ; \ \dot s = \zeta \ ; \ 
\dot \zeta = (p^2/s^3) - (1/s) \ .
$$
The first step is taken with $dt = 0.001$ with the calculation proceeding whenever
the discrepancy [ with subscripts indicating the number of steps ] :
$$
\sqrt{(q_1-q_2)^2 +(p_1-p_2)^2 + (s_1-s_2)^2 + (\zeta_1-\zeta_2)^2 }
$$
is less than $10^{-10}$ and otherwise proceeding with a timestep half as large.
When the discrepancy falls below $10^{-12}$ the timestep is doubled.  As a good
first exercise the reader is encouraged to reproduce {\bf Figure 2} .  For the
relatively short time interval in the Figure the difference between double- and
quadruple-precision simulations (see below) is barely noticeable.

Any number of modifications of the algorithm can be considered and constitute
useful student exercises. There is an interesting downside to existing adaptive
integrators.  Because the {\it past} is different to the {\it future} adaptive
integrators are not reversible.  This means that the ``accuracy'' of such
algorithms cannot be checked directly.  One can easily check a conserved quantity,
like Newtonian energy, equally well in both time directions, but trajectory
accuracy requires a more elaborate investigation\cite{b12}.  One criterion is
the crossing of the $p=0$ section.  One should get the same number of crossings
forward and backward with a good adaptive integrator. 

Because atomistic mechanics problems are typically time-reversible, as are
{\it all} the problems discussed in the present work, we thought it desirable to
develop a {\it time-reversible} adaptive integrator.  We have approached this by
averaging the forward and backward errors at each point in order to define an
error independent of the direction of time, using that error to decide on the
magnitude of the local timestep.  Although this doubles the computational
time it is quite a reasonable price to pay for a robust integrator intended for
demonstration problems with time-reversible dynamics.  Because our approach is
only approximate developing a time-reversible adaptive integrator remains a
worthy research goal.

\begin{figure}
\includegraphics[width=4.5in,angle=+90.]{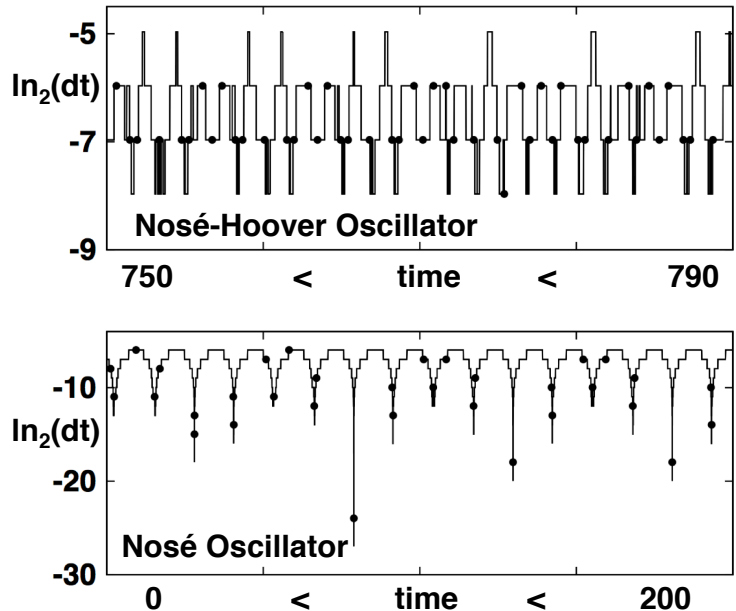}
\caption{
Nos\'e and Nos\'e-Hoover $(q,p,s,\zeta)$ oscillator variation of $\ln_2(dt)$ with
time using double-precision adaptive integration with the rms error confined
between $10^{-12}$ and $10^{-10}$.  Penetrations of the $p=0$ plane are
indicated by filled circles.  The Nos\'e data cover about 88.7 thousand timesteps
while the Nos\'e-Hoover correspond to the range from 80.2 thousand to  84.4
thousand  timesteps.  In both cases the
initial condition is $(q,p,s,\zeta,{\cal H}) = (2.4,0,e^{-2.88},0,0)$ .
}
\end{figure}

The freeware ``gfortran'' FORTRAN compiler from the GNU Project is routinely used
in numerical work with ``double precision'' arithmetic (about sixteen decimal digits)
through the global typing declaration for floating-point variables :\\
{\tt
implicit double precision (a-h,o-z) \\
}
\noindent
With this convention variables in the interval {\tt i-n} are integers.
For stiff problems {\it quadruple precision} simulations (with about thirty-four
decimal digits) are useful.  In gfortran this entails a minor change in the compile
command : \\
{\tt
gfortran -O -o xcode code.f $\rightarrow$
gfortran -O -o xcode -freal-8-real-16  code.f\\
}
 
\begin{figure}
\includegraphics[width=5.5in]{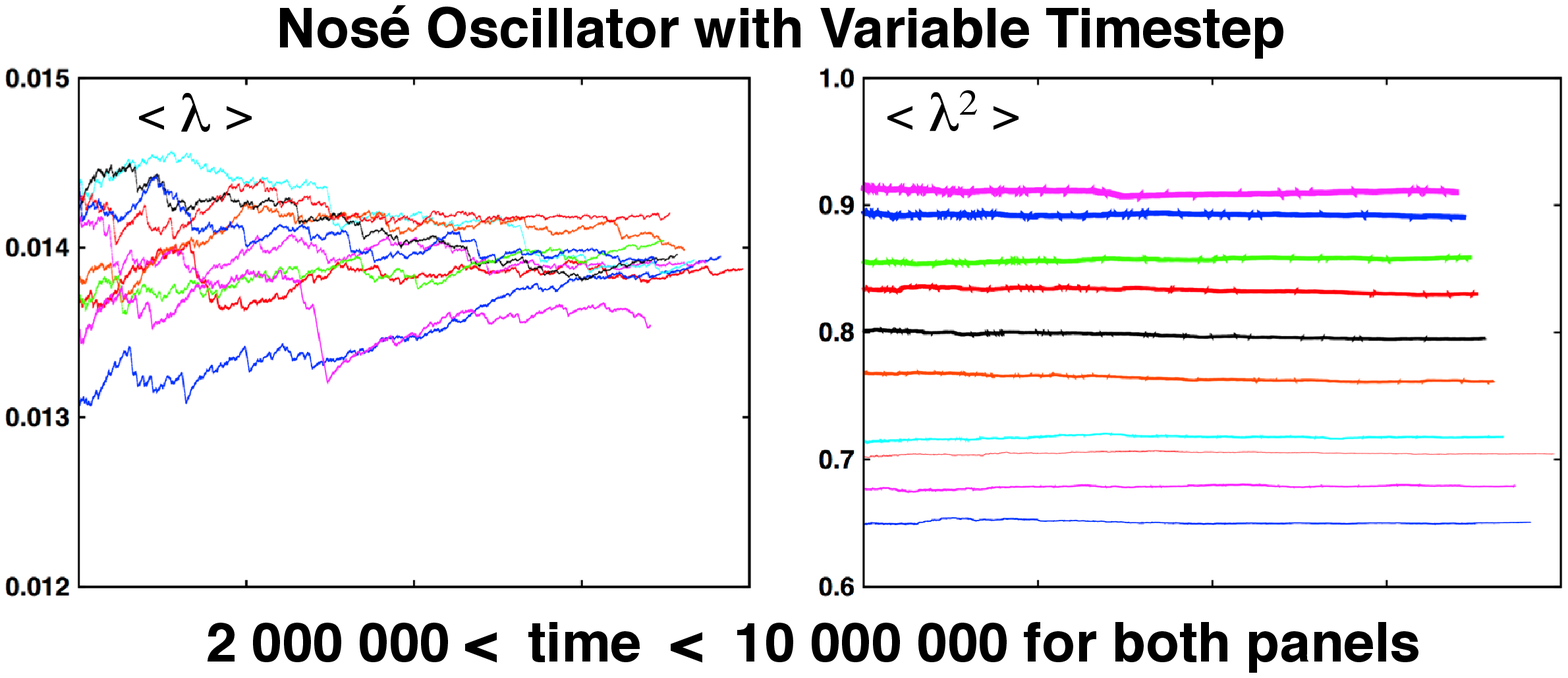}
\caption{
Nos\'e oscillator time variation of $\langle \ \lambda_1 \ \rangle$ and
$\langle \ \lambda_1^2 \ \rangle$ using a reference-to-satellite offset of
$\delta = 10^{-6}$ and double-precision integration. The rms error is confined
within the band from $10^{-12}$ to $10^{-10}$ . As pointed out in the text the
time-averaged Lyapunov exponent is temperature-independent.  On the other hand
the {\it fluctuation} of the exponent varies with temperature as shown in the right
panel.  The ten curves correspond to equally-spaced temperatures from 1 to 10 .
The time reached is a uniformly-decreasing function of temperature.
}
\end{figure}

These oscillator models have both {\it stable} and {\it unstable} ``chaotic''
solutions.  In the latter case the effects of small changes in the initial
conditions increase exponentialy in time as the solution proceeds.
Choosing such a chaotic initial condition for the Nos\'e oscillator and an initial
timestep $dt = 0.001$ generates a billion-timestep adaptive trajectory without any
numerical difficulties.  {\bf Figure 3 } shows the variation of the adaptive timestep
on a semilogarithmic plot for a chaotic problem.
As a check of such sensitive trajectories other integrators
can be used and compared. Though chaos prevents our ever finding {\it the}
solution to an initial-value dynamics problem it cannot prevent our finding a
{\it reasonable} one !

\section{Numerical Studies of the Oscillator Problems}

With our computational tools well in hand, both double- and quadruple-precision,
let us turn to the numerical characterization of the stiffness and chaos in the
oscillator problems.  These results are new and are indications of many new and
promising research directions, from simple exercises to fully-fledged thesis
work.  To begin we quantify the Lyapunov instability of a chaotic trajectory by
{\it measuring} the rate at which two neighboring trajectories diverge.  If 
$\delta$ represents their separation, we characterize its single-timestep tendency
toward divergence by the local Lyapunov exponent,
$\lambda(t) \equiv (\dot \delta/\delta)$ .  {\bf Figure 4} shows that the
dependence of the time-averaged exponent on temperature disappears for long times.
The {\it fluctuation} in $\lambda$, shown to the right, has a significant dependence
on temperature. Let us explain the details of such simulations.
\begin{figure}
\includegraphics[width=3.5in,angle=+90.]{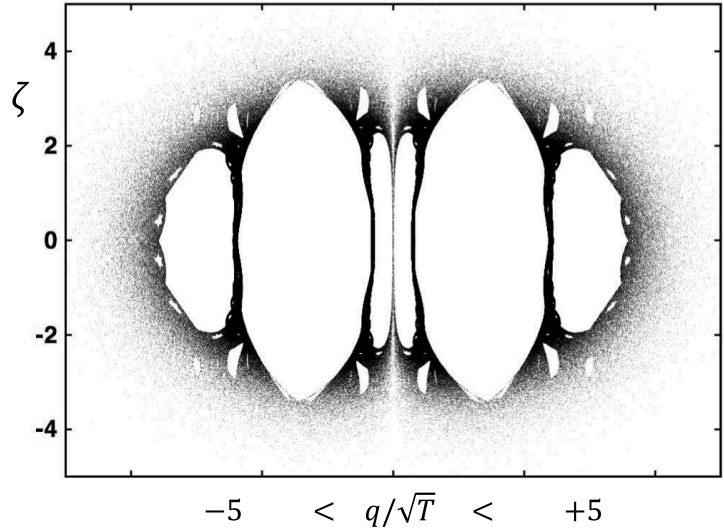}
\caption{
The $(q,p,s,\zeta,{\cal H}) = (q,0,s,\zeta,0)$ cross-section of the Nos\'e oscillator's
isoenergetic three-dimensional chaotic sea.  Stable tori occupy the infinitely-many
holes in the distribution.  See Reference 3 for many examples. The largest Lyapunov
exponent in the sea is $\lambda_1 = 0.046$ so that the chaos is relatively ``weak''
or ``slow''.  {\it Exactly the same} cross-section results from the isothermal
Nos\'e-Hoover oscillator equations, for which the largest Lyapunov exponent
is even slower, 0.0145.  Despite the similar Lyapunov exponents and identical
cross sections the Nos\'e oscillator has huge fluctuations in its rate of
phase-space exploration, and requires an adaptive integrator in order to reproduce
the cross-section shown here.  The abscissa reflects the scaling of the coordinate
$q$ and momentum $p$ [ $(p/s)$ in the Nos\'e-Hoover case ] with the square root of
the temperature $T$ .  This million-point section applies to both the Nos\'e and
Nos\'e-Hoover oscillator trajectories which share a common chaotic sea when ${\cal H}$
vanishes.
}
\end{figure}

\subsection{Chaos and Characterization of the Nos\'e Oscillator's Stiffness}

\begin{figure}
\includegraphics[width=3.5in,angle=+90.]{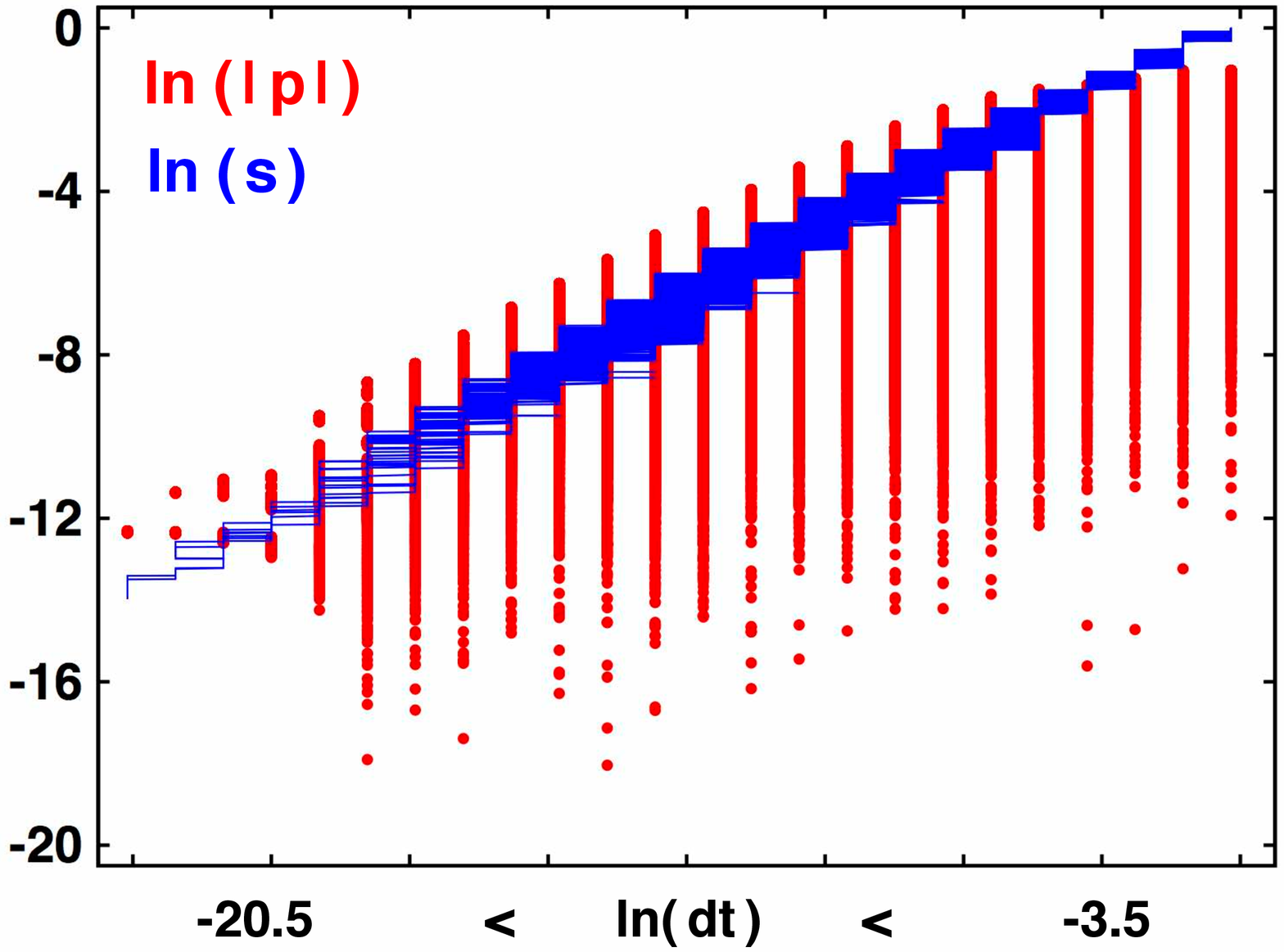}
\caption{
Dependence of the speed $| \ p \ | $ and scale factor $s$ on the variable
timestep $dt$.  Both variations are close to linear.  The adaptive fourth-order
Runge-Kutta integrator maintains a single-step rms error between the limits of
$10^{-12}$ and $10^{-10}$ .  The plot includes one million double-precision Nos\'e
oscillator timesteps in the chaotic sea with ${\cal H} = 0$ .
}
\end{figure}

Nos\'e's oscillator problem has two kinds of solution, [ 1 ] relatively
smooth and stable periodic or toroidal solutions, and  [ 2 ]
{\it unpredictable} and {\it chaotic} and {\it unstable} solutions in
which nearby trajectories separate from one another exponentially
fast, exhibiting ``Lyapunov instability'', $\delta \simeq e^{+\lambda t}$.
{\bf Figure 5} shows a cross-section of the chaotic sea.  The points are
plotted whenever a trajectory passes through the plane $p=0$ or $(p/s)=0$.
They therefore reflect the product of the probability density and the speed
normal to the plane, $| \ \dot p \ |$ , the ``flux'' :
$$
| \ q \ |f(q,0,s,\zeta) \propto | \ q \ |f(q,0,\zeta) \propto
| \ q \ |e^{-q^2/2}e^{-\zeta^2/2} \ .
$$
The stationary distribution satisfying $(\partial f/\partial t) = 0$ is
Gaussian in all three Nos\'e-Hoover state variables $(q,p,\zeta)$ . The
cross section in {\bf Figure 5}, and even the flux through it, are exactly
the same in Nos\'e and Nos\'e-Hoover dynamics ( because the trajectory
is the same, with the same velocity {\it at} the $p=0$ plane, independent
of $s$ ) :
$\dot p_N = \dot p_{NH} = -q$ .

Let us delve into the details of the chaotic sea from the perspective of
Nos\'e's oscillator, using what we consistently adopt as our standard
chaotic initial condition :
$$
(q,p,s,\zeta) = (2.4,0,e^{-2.88},0) \longrightarrow {\cal H}_N = (1/2)
[ \ q^2 + (p/s)^2 + \ln(s^2) + \zeta^2 \ ]  \equiv 0 \ .
$$
One way ( there is no consensus ) to quantify the oscillator's ``stiffness'' is
to record the range of the time-scaling factor $s$ which is responsible for the
stiffness.  Another way is to record the range over which the timestep must be
varied in order to solve the equations with a given integrator.  We have used
the classic fourth-order Runge-Kutta integrator, increasing or decreasing $dt$
as needed for accuracy.  All three algorithmic ``variables'' $(s,| \ p \ |,dt)$
show a roughly linear correlation in the log-log plot of {\bf Figure 6}.
The stiffness gives a rough correspondence between $\ln(dt)$ and $\ln(s)$ .
Because the scaled momentum is proportional to $\sqrt{T}$ , one in the figure,
$| \ p \ |$ is roughly proportional to $s$ and to $dt$ . To solve this same
Nos\'e oscillator problem with a {\it fixed} timestep would require
$dt \simeq 0.0000001$ .

By way of contrast the smooth Nos\'e-Hoover \#2 equations can be solved with
the classic fourth-order Runge-Kuttta integrator for two billion double-precision
timesteps (roughly 15 decimal digits) with $dt = 0.01$ ,  reaching a time of
20,000,000 without any difficulty. The three-dimensional initial condition,
$(q,p,\zeta) = (2.4,0,0)$ , corresponds to the four-dimensional $(q,p,s,\zeta)
= (2.4,0,e^{-2.88},0)$
condition used in {\bf Figure 2}.  The Nos\'e-Hoover solution generates more than
five million penetrations of the $p=0$ cross-sectional plane.  With this fixed
timestep the Nos\'e oscillator's progress is roughly one hundred thousand times
slower, while generating {\it exactly the same} $(q,p/s,\zeta)$ states.

\subsection{Chaos and the Local Lyapunov Exponent $\lambda(t)$ as a Stiffness Criterion}

An alternative measure of stiffness can be based on the local ( instantaneous )
Lyapunov exponent.  The local exponent describes the rate $\lambda(t)$ at which
two nearby trajectories tend to separate, $ \dot \delta = \lambda \delta $ .
Choosing a satellite trajectory $x_s$ constrained to a distance $\delta =
0.000001$ from the reference $x_r$ , the distance is rescaled after a time $dt$
by multiplying the separation by a factor $g$ ( which is close to unity ) :
$$
x_s = x_r + g(x_s - x_r) \ {\rm where} \ g =
( \ \delta/\sqrt{(x_s-x_r)^2} \ ) \simeq e^{-\lambda dt} .
$$
The local Lyapunov exponent is $\lambda(t) = -\ln(g)/dt \ $.  {\bf Figure 7}
shows two sections of a typical time history of $\lambda$ from an adaptive
solution of Nos\'e dynamics. The central Nos\'e peak at a time of 5.0862 was
resolved by an adaptive integrator which chose a timestep of $2 \times 10^{-9}$
for the peak shown.  The corresponding time for Nos\'e-Hoover \#2 dynamics is
16.42, and corresponds to a broad minimum in $s$. The amplitudes and the required
numbers of timesteps differ by about six orders of magnitude for the two equivalent
representations of a thermostated oscillator trajectory.

\begin{figure}
\includegraphics[width=4.5in,angle=+90.]{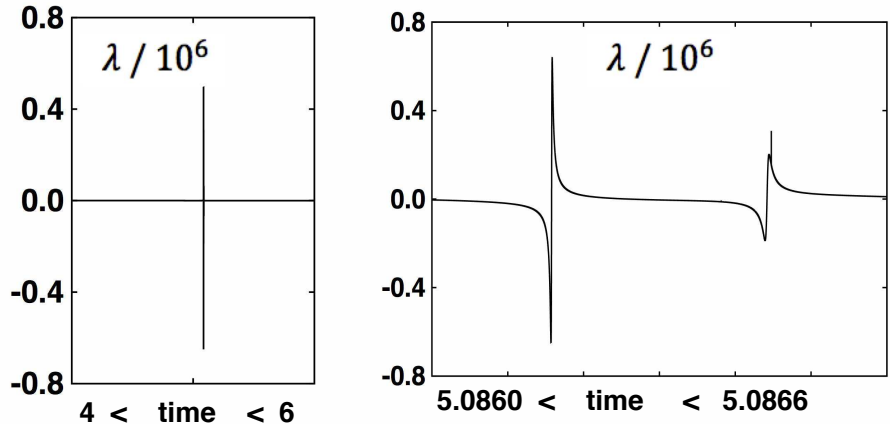}
\caption{
Nos\'e oscillator local Lyapunov exponent in the chaotic sea as a function of
time with $\delta = 10^{-6}$ and the quadruple-precision error confined to the
band between $10^{-28} \ {\rm and} \ 10^{-24}$.  The left panel shows data with
timesteps from 1.63 to 3.07 million.  The right panel includes data between
2.01 and 2.16 million timesteps.  The standard chaotic initial condition
$(q,p,s,\zeta,{\cal H}) = (2.4,0,e^{-2.88},0,0)$ .
}
\end{figure}

\begin{figure}
\includegraphics[width=4.5in,angle=+90.]{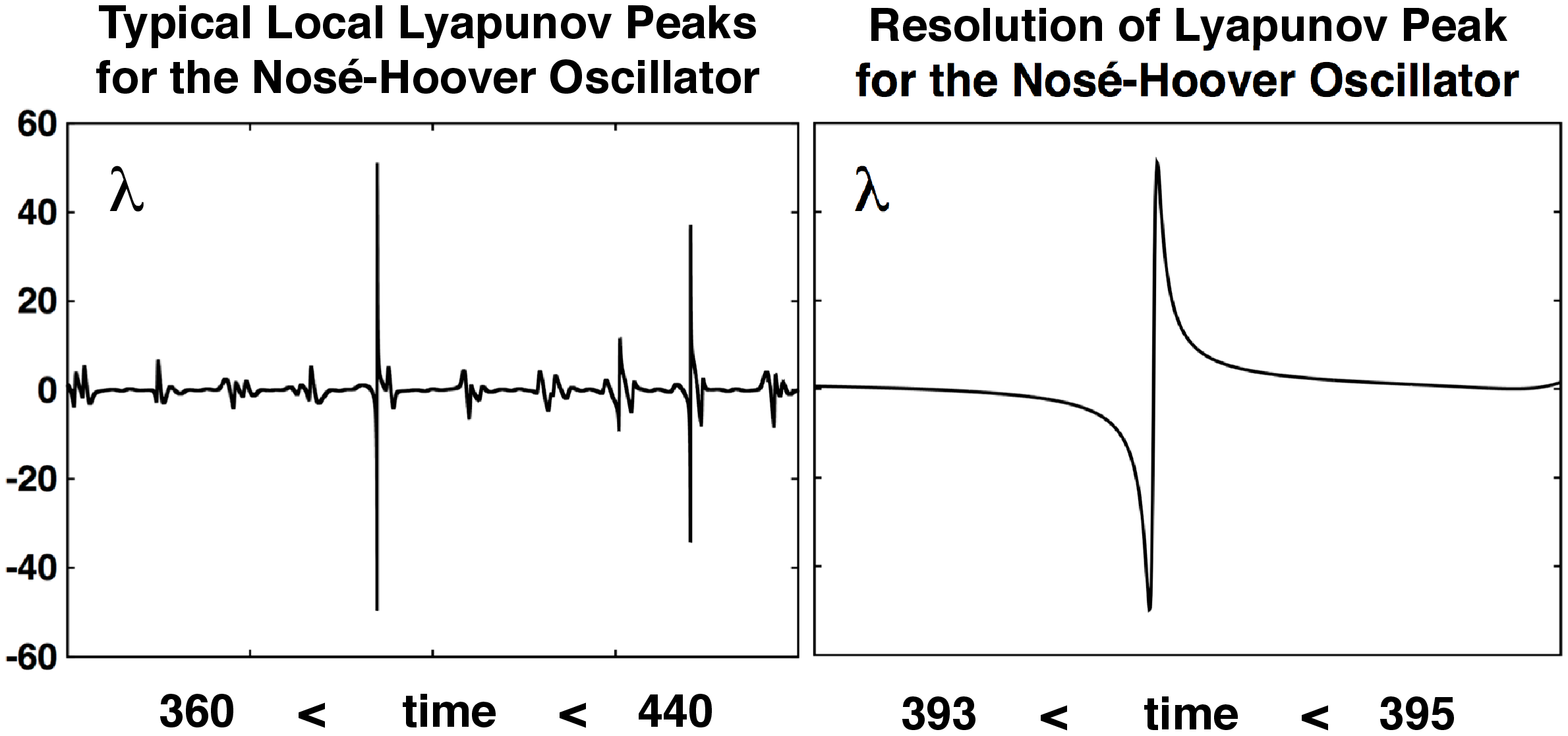}
\caption{
Nos\'e-Hoover $(q,p,s,\zeta)$ local Lyapunov exponent variation in the chaotic
sea with a closeup showing the resolution of a typical ``large'' peak.  The
maximum time shown, 440, corresponds to a total of 286 575 timesteps with $dt$
confined to the interval $10^{-16}$ to $10^{-14}$ using double-precision
arithmetic.  The satellite-to-reference separation is $\delta = 10^{-6}$ .
}
\end{figure}

Nos\'e got around the stiffness of his motion equations by arbitrarily
multiplying the righthand sides of each of his four Hamiltonian equations of
motion by $s$.  This trick doesn't change the four-dimensional trajectory at
all if we visualize the trajectory as a one-dimensional path in four-dimensional
$(q,p,s,\zeta)$ space. But the {\it rate} at which the path is followed is
changed by the factor $s$.  Nos\'e termed this change of rate ``scaling the
time" and uses it in his 1984 papers. His ``real'' {\it versus}  ``virtual''
variables helped to make his work relatively difficult to fathom.  Now that we
have been able to solve his equations directly, with adaptive integrators we
have attained a good picture of the Lyapunov instability and stiffness of the
original Nos\'e equations.  Let us summarize the present situation.

\section{Summary and Outlook}

\subsection{Adaptive Integrators and the Use of Mappings}

Today's ``realistic'' computer models, typically representing thermostated aqueous
solutions, involve a host of practical computational issues.\cite{b13}  One
we did not detail here is the need for integration techniques dealing with
discontinuous righthand sides. Adaptive integrators are often used in artificial
demonstration problems with {\it singular} righthand sides. 
Discontinuous jumps of a control variable [ ``Bang-Bang Control'' ] is an example.

Special  precautions need to be taken when the dynamics itself is singular ( as in
hard-sphere or square-well dynamics ), replacing the smooth separation rate
$\lambda(t)$ with a singular {\it map}\cite{b14} as the collisional ``events''
occur.  Although this has been the typical approach in molecular dynamics
simulations since the 1980s, the mapping technique is unfamiliar to most workers
in dynamical systems simulations. In hard-sphere molecular dynamics, going back
to Alder and Wainwright's pioneering work of the middle 1950s\cite{b15}, with
thousands or millions of degrees of freedom, it is usual to integrate up to the
moment of the next collision, change the momenta of the colliding particles
{\it at} the collision, and then continue on until the next collision.  These
``event-driven'' simulations are another example of the need for two or more
solution strategies for the underlying differential equations.

\subsection{Characterizing Chaos with Lyapunov Instability}

Where chaos is concerned, which is typical of ``interesting'' problems, Lyapunov
instability is most easily quantified by following two neighboring trajectories.
The relative motion of the two trajectories can be constrained by using a Lagrange
multiplier. Alternatively the distance between the two trajectories can be rescaled.
Either way the original separation length is recovered at the end of each timestep.
The logarithm of the scale factor required to do this is simply related to the
local Lyapunov exponent.  These two approaches provide the largest Lyapunov exponent
$\lambda_1$ , the mean separation rate over a long simulation,
$\dot \delta = \lambda_1 \delta$ .  The Lagrange multiplier approach
can also be applied to a linearized version of the equations at the expense of
additional algebra.  All three of these techniques are useful tools which can
benefit from adaptive integration\cite{b16}. 

Chaotic systems will always be a challenge.
Joseph Ford\cite{b17} emphasized that numerical methods are unable to follow any chaotic
trajectory accurately for very long.  The only convincing test of accuracy is
the reproducibility of the trajectory itself.  Simple reversal of a trajectory
or conservation of energy are not reliable criteria for accuracy\cite{b15}.
Numerical chaotic trajectories can simulate these difficult situations and
do  provide ``weak'' averages which are accurate despite the lack of global
accuracy in the computed trajectory. This good fortune is likewise typical
of event-driven dynamical systems.  It is evident that {\it no} numerical
methods are capable of precise solutions of such problems.  Adaptive
integrators are a useful tool for producing ``reasonable'' chaotic trajectories,
inaccurate though they may be.

\subsection{Unique Features of the Nos\'e-Based Oscillator Models}

\begin{figure}
\includegraphics[width=5.5in,angle=+90.]{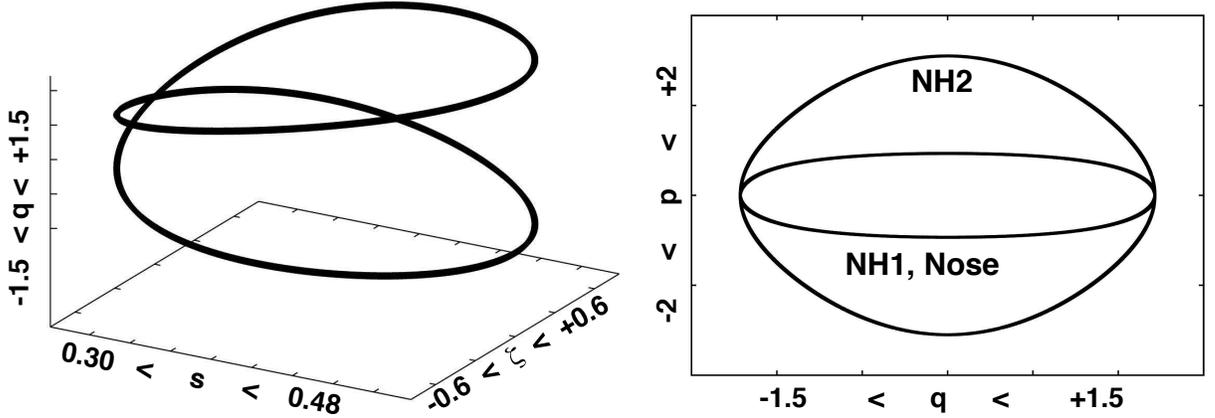}
\caption{
At the left the vertical coordinate is $q$,  plotted as a function of $s$, which is
always positive, and $\zeta$ .  At the right the smaller ellipse corresponds to
to both the Nos\'e and Nos\'e-Hoover \#1 equations.  The outer ellipse represents
Nos\'e-Hoover \#2.  The two Nos\'e-Hoover curves have a period of 5.578 while the
Nos\'e period is ``faster'' 2.1655 .
}
\end{figure}

Because all three oscillator models, [ Nos\'e, Nos\'e-Hoover \#1, and Nos\'e-Hoover \#2,
or $N$ and $NH$ ] share common trajectories, their travel times and their Lyapunov
exponents are related in an intriguing way, unique in our dynamical-systems experience.
At unit temperature the three models are precisely related: $\lambda_N = 3.26\lambda_{NH}$
and $\vec{t}_{NH} = 3.26\vec{t}_N$. Here the $\{ \ \vec{t} \ \}$ denote travel times
along a sufficiently long trajectory.  The Nos\'e trajectory is the fastest of the three.
The disparities are the result of the slower rates (by a common factor $s<1$) in the
two versions of Nos\'e-Hoover dynamics. The travel times for a long trajectory in
$(q,p,s,\zeta)$ space are necessarily related by the same ratio, with
$\vec{t}_{NH} = 3.26\vec{t}_N$ . The longtime averaged growth rate of the slower
Nos\'e-Hoover trajectories leads to a smaller common Lyapunov exponent
$\lambda_{NH\#1} = \lambda_{NH\#2} = \lambda_N/3.26$ .  This relationship
between the time-averaged exponents does {\it not} hold for the local exponents,
$\lambda_{NH}(t)$ and $\lambda_N(t)$ .

Why not? It is evident that both dynamics share the {\it same} reference trajectory.
It might then appear that the neighboring satellite trajectory determining $\lambda(t)$
would be the same for both dynamics too.  But because the local exponent responds to the
time-rate-of-change of the scale factor $s(t)$ there is no simple relationship linking
the local exponents.  Nevertheless the relation between the rates,
$(d/dt)_{NH} \equiv s(d/dt)_N$, is valid for any $(q,p,s,\zeta)$ trajectory segment with
a vanishing Hamiltonian.  The relation linking all the rates leads directly to the useful
relations :
$$
\vec{t}_{NH} = \langle \ 1/s \ \rangle_N \vec{t}_N \ ; \
\vec{t}_N = \langle \ s \ \rangle_{NH}\vec{t}_{NH} \ .
$$

As an example, illustrated in {\bf Figure 9}, consider the simplest stable periodic orbit,
with initial values $(q,p,s,\zeta,{\cal H}) = (1.2145,0.0,e^{-q^2/2},0,0)$ , traced out
in times of $\vec{t}_{NH} = 5.5781$ and $\vec{t}_N = 2.1655$ .  The ratios of the two times
are related to the mean values of $s^{\pm 1}$ averaged over the orbit :
$$
\vec{t}_{NH}/\vec{t}_N = 5.5781/2.1655 = 2.5759 = \langle \ s^{-1} \ \rangle_N \ ;
$$
$$
\vec{t}_N/\vec{t}_{NH} = 2.1655/5.5781 = 0.3882 = \langle \ s^{+1} \ \rangle_{NH} \ .
$$

An interesting feature of the oscillator problems is that {\it the largest
Lyapunov exponent is independent of temperature}.  See the lefthand side of
{\bf Figure 4}.  The righthand side {\it does} show temperature dependence. The
{\it fluctuation} of the exponents
( hardly distinguishable from $\langle \ \lambda^2_1(t) \ \rangle$ ) increases
with temperature for temperatures greater than 2.

We can {\it prove} the temperature independence shown in {\bf Figure 4}.  If we consider the
{\it temperature-dependent} Nos\'e-Hoover \#2 equations and introduce new variables, 
$$
Q \equiv (q/\sqrt{T}) \  ; \ P \equiv (p/\sqrt{T}) \ ; \ Z = \zeta \ , 
$$
the temperature-independent result shows that the largest Lyapunov exponent
( but not its fluctuation ) is indeed temperature independent ( This is because
a {\it linear} change of scale leaves the time-averaged {\it logarithmic} growth
rate unchanged ) :
$$
\dot q = p \rightarrow \dot Q = P \ ; \
\dot p = - q - \zeta p \rightarrow \dot P = -Q - Z P \ ; \
\dot \zeta = (p^2/T) - 1 \rightarrow \dot Z = P^2 - 1 \ .
$$

In Section IIB we saw that the temperature-dependent $(q,p,\zeta,T)$
Nos\'e-Hoover equations provide a stationary solution from Liouville's Theorem :
$$
f \propto e^{-q^2/2T}e^{-p^2/2T}e^{-\zeta^2\tau^2/2} \rightarrow \
$$
$$
\dot q = p \ ; \ \dot p = -q - \zeta p \ ; \
\dot \zeta = [ \ (p^2/T) - 1 \ ]/\tau^2 \ ,
$$
Just as in the simpler case with $T$ and $\tau$ equal to unity we can find 
a corresponding ${\cal H}_D$ :
$$
2{\cal H}_D = sq^2 + (p^2/s) + Ts\ln(s^2) + (s\zeta^2/T\tau^2) \equiv 0 \ .
$$

We remarked on the disparity of the Nos\'e and Nos\'e-Hoover Lyapunov exponents.
The Nos\'e oscillator exponent is $0.046\pm 0.001$ while the two Nos\'e-Hoover
exponents are smaller, $0.014_5 \pm 0.0001$ .
{\bf Figure 7} illustrated, on two very different scales, a typical
excursion of the local Lyapunov exponent in Nos\'e's stiff case.  On the
scale of the figure at the left the variation looks {\it singular}, but it
is actually {\it smooth} when adaptive integration is used to control the
scale of the smallest timestep, which is $10^{-9}$ .  {\bf Figure 8} shows
similar data for the Nos\'e-Hoover version of the same problem.  In
that case no special precautions need be taken.  A fixed timestep of
0.001 is perfectly adequate for accurate estimates of the local Lyapunov
exponent.

\subsection{Challenges and Ideas for Future Work}

In bringing this discussion of adaptive integrators to a close let us mention
that there is a significant variation in the {\it topology of} the chaotic and
regular solutions for our oscillator models where the thermostating is ``fast'' :
$$
\dot \zeta = [ \ p^2 - 1 \ ] \longrightarrow \dot \zeta = [ \ p^2 - 1 \ ]/\tau^2
$$
[ when the response time $\tau$ is small ]. In particular, fast thermostating
generates infinite numbers of  intricately {\it knotted trajectories}, many of
which are described in References 18 and 19 . These simple oscillator models,
with quadratic ordinary differential equations (and their nonequilibrium fractal
relatives, where temperature is a function of the coordinates), can generate
{\it interlinked rings} in phase space.  An example is described in our own very
recent work with Puneet Patra\cite{b20}.  These models' chaotic trajectories,
with their knots, and interlinked rings could easily fill an entertaining
and profusely-illustrated Book on the subject.  We urge the reader to explore
and enjoy these topics.  In this spirit we append here three colored Nos\'e $p=0$ sections
illustrating the dependence of $\lambda(t)$, $dt$, and $s$ on $(q,\zeta)$ .

\begin{figure}
\includegraphics[width=4.5in,angle=+90.]{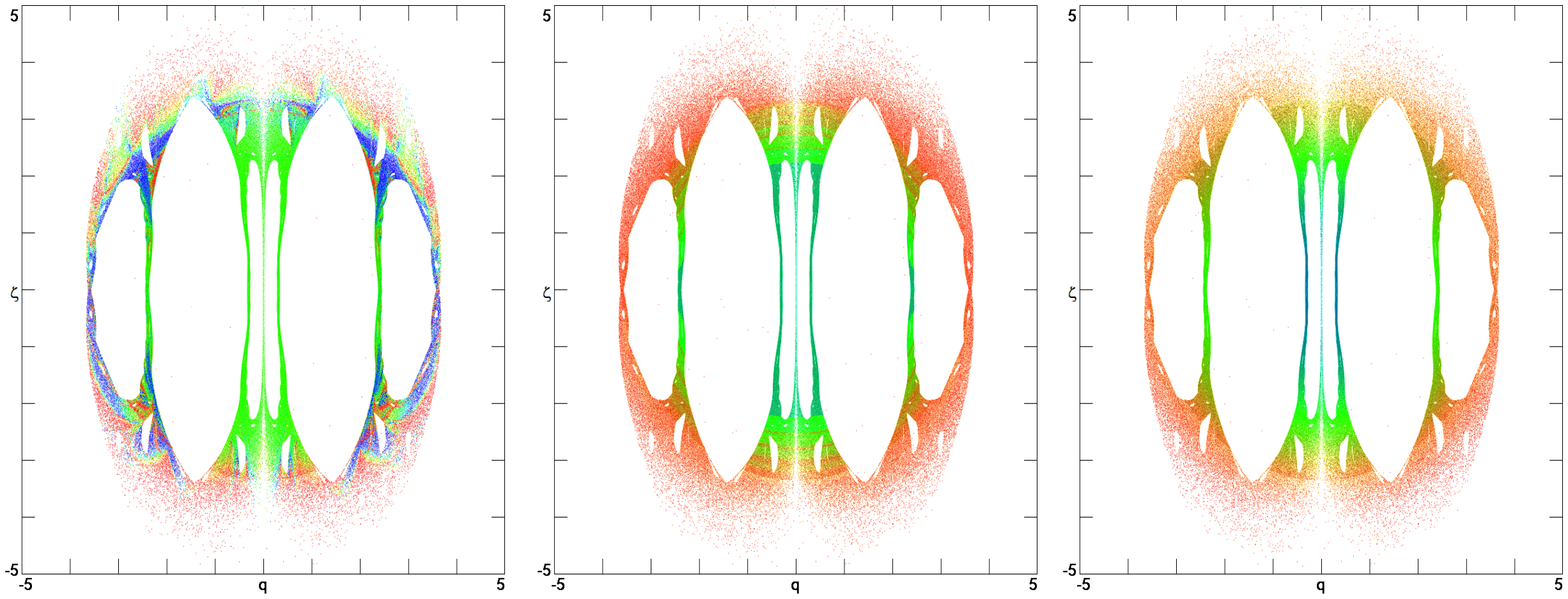}
\end{figure}

\pagebreak

\end{document}